\documentclass[a4paper,english,prl,reprint,twocolumn,showpacs,superscriptaddress,groupedaddress]{revtex4}
\usepackage[T1]{fontenc}
\usepackage[latin9]{inputenc}
\usepackage{babel}
\usepackage{textcomp}
\usepackage{amstext}
\usepackage{amssymb}
\usepackage{graphicx}
\usepackage[unicode=true,pdfusetitle,
 bookmarks=true,bookmarksnumbered=false,bookmarksopen=false,
 breaklinks=false,pdfborder={0 0 1},backref=false,colorlinks=false]
 {hyperref}

\makeatletter

\pdfpageheight\paperheight
\pdfpagewidth\paperwidth

\DeclareRobustCommand{\greektext}{%
  \fontencoding{LGR}\selectfont\def\encodingdefault{LGR}}
\DeclareRobustCommand{\textgreek}[1]{\leavevmode{\greektext #1}}
\DeclareFontEncoding{LGR}{}{}
\DeclareTextSymbol{\~}{LGR}{126}
\newcommand{\lyxmathsym}[1]{\ifmmode\begingroup\def\b@ld{bold}
  \text{\ifx\math@version\b@ld\bfseries\fi#1}\endgroup\else#1\fi}

\@ifundefined{textcolor}{}
{%
 \definecolor{BLACK}{gray}{0}
 \definecolor{WHITE}{gray}{1}
 \definecolor{RED}{rgb}{1,0,0}
 \definecolor{GREEN}{rgb}{0,1,0}
 \definecolor{BLUE}{rgb}{0,0,1}
 \definecolor{CYAN}{cmyk}{1,0,0,0}
 \definecolor{MAGENTA}{cmyk}{0,1,0,0}
 \definecolor{YELLOW}{cmyk}{0,0,1,0}
 }

\makeatother

\begin{document}

\title{Fiber transport of spatially entangled photons}

\author{W. Löffler}

\email{loeffler@physics.leidenuniv.nl}

\affiliation{Huygens Laboratory, Leiden University, P.O. Box 9504, 2300 RA Leiden,
The Netherlands}

\author{T. G. Euser}

\affiliation{Max Planck Institute for the Science of Light, Günther-Scharowsky-Str.
1, 91058 Erlangen, Germany}

\author{E. R. Eliel}

\affiliation{Huygens Laboratory, Leiden University, P.O. Box 9504, 2300 RA Leiden,
The Netherlands}

\author{M. Scharrer}

\affiliation{Max Planck Institute for the Science of Light, Günther-Scharowsky-Str.
1, 91058 Erlangen, Germany}

\author{P. St.J. Russell}

\affiliation{Max Planck Institute for the Science of Light, Günther-Scharowsky-Str.
1, 91058 Erlangen, Germany}

\author{J. P. Woerdman}

\affiliation{Huygens Laboratory, Leiden University, P.O. Box 9504, 2300 RA Leiden,
The Netherlands}
\begin{abstract}
Entanglement in the spatial degrees of freedom of photons is an interesting
resource for quantum information. For practical distribution of such
entangled photons it is desireable to use an optical fiber, which
in this case has to support multiple transverse modes. Here we report
the use of a hollow-core photonic crystal fiber to transport spatially
entangled qubits.
\end{abstract}
\maketitle
The non-classical correlations of two entangled photons enable quantum
communication and cryptography. Traditionally, polarization entanglement
is utilized; however, polarization is a two-dimensional degree of
freedom only (qubit). Higher-dimensional entangled systems show promise
\cite{collins2002,fujiwara2003}, because more entangled degrees of
freedom per particle imply stronger non-classical correlations. Examples
are the violation of a Bell-type inequality \cite{vaziri2002,thew2004}
and demonstration of the quantum coin tossing protocol \cite{molinaterriza2005}.
The photon's temporal, frequency and spatial degrees of freedom give
access to such high-dimensional entanglement; enabling temporal entanglement
\cite{franson1989,riedmatten2002,thew2004}, frequency entanglement
\cite{ramelow2009,olislager2010}, and spatial entanglement \cite{rarity1990,mair2001,vaziri2002,langford2004,osullivanhale2005,oemrawsingh2005,jack2009,taguchi2009},
respectively. These degrees of freedom are intrinsically continuous;
after appropriate discretization, qudits of arbitrary dimension $d$
can be defined. Another approach for multi-dimensional entanglement
is the combination of different degrees of freedom \cite{strekalov1996,barreiro2005,yang2005,cinelli2005}.

Optical-fiber transport of entangled photons allows new possibilities
in applied quantum information; so far this has been realized for
photons entangled in polarization \cite{poppe2004}, time \cite{marcikic2004}
and frequency \cite{ramelow2009,olislager2010}, but not for spatial
entanglement. To transport spatially entangled photons through a fiber,
this must obviously support multiple transverse modes. However, this
presents a challenge since conventional multimode fibers suffer from
strong intermodal coupling which tends to destroy the fragile quantum
correlations carried by the spatially entangled state. Often this
happens within a few mm of propagation \cite{lucesoli2007}%
\footnote{Also our own experiments with conventional multi-mode fibers show
decoherence after a few millimeters.%
}. It is thus imperative to choose a type of fiber where these decohering
effects are minimal or absent, i.e., to be as close as possible to
free-space propagation. Therefore, the natural choice for fiber transport
of spatially entangled photons is to use a \emph{hollow-core} fiber,
where the light is guided essentially in air (see inset Fig. 1b).
Here we demonstrate that this approach is successful.

We use a hollow-core photonic crystal fiber (HC-PCF) \cite{knight1998,shapira2010}
with a kagomé-style cladding \cite{benabid2002,couny2006,euser2008},
see Fig.~1b. The cladding lattice is formed from sub-\textgreek{m}m
thick fused silica glass membranes, the central hollow core being
created by removing a few unit cells. Our HC-PCF has a relatively
large core diameter, namely 25 \textgreek{m}m, and does not support
bound modes because the cladding does not possess a photonic band
gap and the negative index difference between core and cladding rules
out the possibility of total internal reflection. Many Mie-like resonances
exist in the core, each with a different axial wavevector component.
In some respects these modes are similar to the leaky modes of hollow
cylindrical capillary waveguides \cite{marcatili1964}, with an important
difference: with appropriate design the fundamental mode of the HC-PCF
can be guided with losses as much as 100 times lower than in a capillary
of the same diameter \cite{couny2006}. It has a transverse intensity
profile that approximately follows a squared $J_{0}$ Bessel function,
its first zero coinciding with the core boundary. The losses are larger
for higher-order modes and this limits effectively the number of propagating
modes. In the experiments reported here, 3 modes are present after
propagation along 30 cm of our kagomé HC-PCF. These modes have large
overlap with the 3 lowest-order free-space paraxial Hermite-Gaussian
modes (or superpositions thereof). In the Hermite-Gauss basis $HG_{m,n}$
with the polynomial index $m$ and $n$ in $x$ and $y$, these are
the $HG_{0,0}$, $HG_{1,0}$, and $HG_{0,1}$ modes. We find experimentally
that the 30~cm kagome fiber attenuates the first order modes by $\approx$8\%
compared to the fundamental mode, this agrees well with 0.7 dB/m or
5\% as quoted in \cite{euser2008} for a similar fiber. 

\begin{figure}
\includegraphics[width=1\columnwidth]{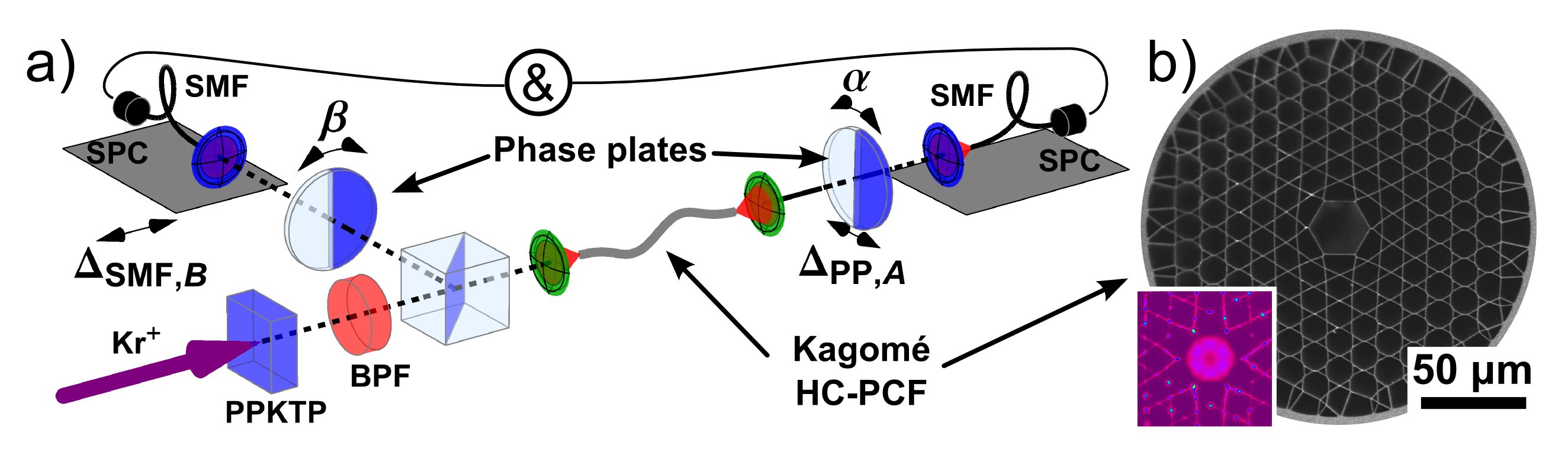}

\caption{(Color online) (a) Experimental setup. (b) Electron micrograph of
the cross-section of the HC-PCF. The core diameter is 25~$\lyxmathsym{\textgreek{m}}$m,
which results in an half opening angle of the fundamental mode of
44~mrad. The inset shows a near-field optical image (same scale)
of the core area, acquired using incoherent (800\textpm{}10~nm) illumination,
this shows approximately circular mode confinement.}
\end{figure}

In our experiments we study the effect of fiber transport on quantum
correlations between two entangled qubits, where spatially entangled
photons are produced by spontaneous parametric downconversion (SPDC,
type-I, collinear phase matching) \cite{mair2001}. The setup is sketched
in Fig.~1: The entangled two-photon state is generated in a 2~mm
long periodically-poled KTP crystal by pumping it with a weakly focused
($w_{0}=250$~\textgreek{m}m) 413~nm Kr$^{+}$ laser beam (80~mW).
The polarization state of the photons is not relevant here since polarization
and spatial degrees of freedom are decoupled (paraxial regime). The
two entangled photons are filtered spectrally (BPF, 1~nm bandwidth,
centered at 826.1~nm), then the photons are probabilistically separated
by a beam splitter, and one of them (path A) is transported through
the 30~cm long HC-PCF by appropriately mode-matched in- and out-coupling
(10x, 0.25~NA objectives). For our goal it is sufficient that only
one photon passes through the HC-PCF. No special care was taken to
keep the fiber straight; typically curves with a bending radius of
30~cm occurred. Both photons from a pair are then projected onto
separate superpositions. This is done by first sending the photons
through a step phase plate \cite{pors2008}; this plate acts as a
mode converter; it shifts the optical phase by $\pi$ in one half
of the transverse plane with respect to the other half. Subsequently,
we project this mode onto the fundamental Gaussian mode by coupling
into a standard single-mode fiber. This fiber is connected to a single-photon
counter and correlated photon pairs are post-selected by coincidence
detection (2 ns time-window). The entangled two-photon state as produced
by SPDC \cite{mair2001,walborn2005,pors2008} is filtered by the {}``3-mode''
HC-PCF such that we deal effectively with a 3D spatially entangled
state formed by the 3 lowest-order Gaussian modes \cite{vaziri2002,langford2004}.
The single-photon basis states for these modes are $\left\{ \mid\!\! HG_{0,0}\rangle,\mid\!\! HG_{1,0}\rangle,\mid\!\! HG_{0,1}\rangle\right\} $
in the Hermite-Gauss basis. Such a bipartite entangled state lives
in a $\mathbb{C}^{3}\mathbb{\otimes}\mathbb{C}^{3}$ Hilbert space
which can be explored by investigating spatial correlations in two
non-identical but partly overlapping $\mathbb{C}^{2}\mathbb{\otimes}\mathbb{C}^{2}$
subspaces. 
\begin{figure}
\includegraphics[width=1\columnwidth]{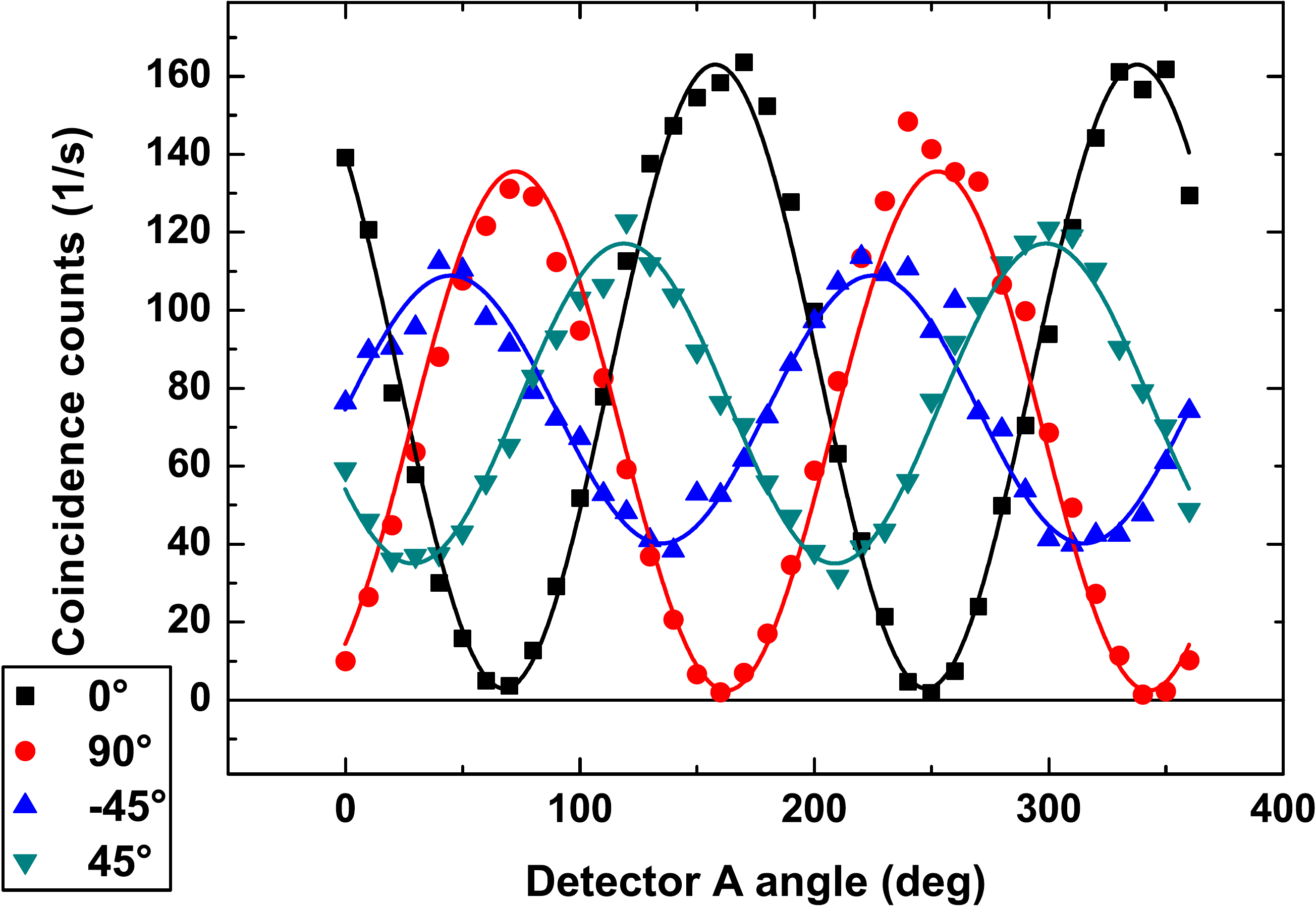}

\caption{(Color online) Fiber transport of 2D spatially entangled photons in
the degenerate case (i.e. superpositions of the $\mid\!\! HG_{1,0}\rangle$
and $\mid\!\! HG_{0,1}\rangle$ states). Both mode analyzers are equipped
with a step phase plate which can be rotated. To obtain a coincidence
fringe, plate A is rotated, while plate B is kept fixed. No significant
fringe is detected in the single-detector count rates. }
\end{figure}

As a first step, we consider the subspace spanned by the \emph{degenerate}
$\mid\!\! HG_{1,0}\rangle$ and $\mid\!\! HG_{0,1}\rangle$ modes.
We center both phase plate projectors to the single-mode detection
fibers ($\Delta_{SMF,B}=\Delta_{PP,A}=0$, see Fig. 1). In the corresponding
2D Hilbert subspace, each detector projects on the superposition state
$\sin(\phi)\mid\!\! HG_{1,0}\rangle+\cos(\phi)\mid\!\! HG{}_{0,1}\rangle$,
which depends on the orientation $\phi$ of the phase plate ($=\alpha$
or $\lyxmathsym{\textgreek{b}}$ for phase plate A or B, respectively,
see Fig.~1a). This experiment is analogous to the 2D polarization-entanglement
case \cite{kwiat1995,padgett1999} and thus, the resulting coincidence
fringes are sinusoidal (Fig.~2). We also see anisotropic effects
of the fiber: the visibility of the diagonal ($\pm45^{\circ}$) fringes
is reduced compared to that of the ($0^{\circ}$, $90^{\circ}$ degrees)
fringes. Also, mode rotation occurs as the fringes are slightly dephased
with respect to the relative orientation ($\alpha-\beta$) of the
phase plates. We attribute these effects to the broken rotational
symmetry in the fiber: the cladding has a hexagonal symmetry; this
weakly affects the modes since a small fraction of the optical field
resides in the cladding. This leads to mode mixing and mode rotation;
both could be avoided if one were to use a circular-core concentric
Bragg fiber \cite{johnson2001}. 

In polarization-based Bell experiments, maximum violation happens
for, e.g., $(\alpha_{1},\alpha_{2},\beta_{1},\beta_{2})=(0^{\circ},45^{\circ},22.5^{\circ},67.5^{\circ})$.
In our case, we expect these angles to be different due to the detrimental
effects mentioned above. In order to find the maximum Bell-violation
angles, we plot the CHSH S-parameter \cite{clauser1969} for $(\alpha_{1},\alpha_{2})=(0^{\circ},45^{\circ})$
as a function of $(\beta_{1},\beta_{2})$ (see Fig.~3). Each pixel
shows color-coded the value of the S-parameter for one set of angles;
regions with $S\!>\!2$ are colored white. The overall pattern of
Fig.~3 is very similar to the ideal case; but the peaks are shifted.
A numerically search for the maximum Bell violation results in the
angles $(\alpha_{1},\alpha_{2},\beta_{1},\beta_{2})=(0^{\circ},-45^{\circ},270^{\circ},150^{\circ})$,
where $S=2.17\pm0.04$. The uncertainty (standard deviation) of $S$
is calculated from the uncertainty $\Delta N$ in the coincidence
counts $N$ using Gaussian propagation of uncertainty. For this 2D
subspace we violate the Bell inequality by 4 standard deviations;
thus proving that spatial entanglement of the degenerate $\mid\!\! HG_{1,0}\rangle$
and $\mid\!\! HG_{0,1}\rangle$ states survives fiber propagation.

\begin{figure}
\includegraphics[width=1\columnwidth]{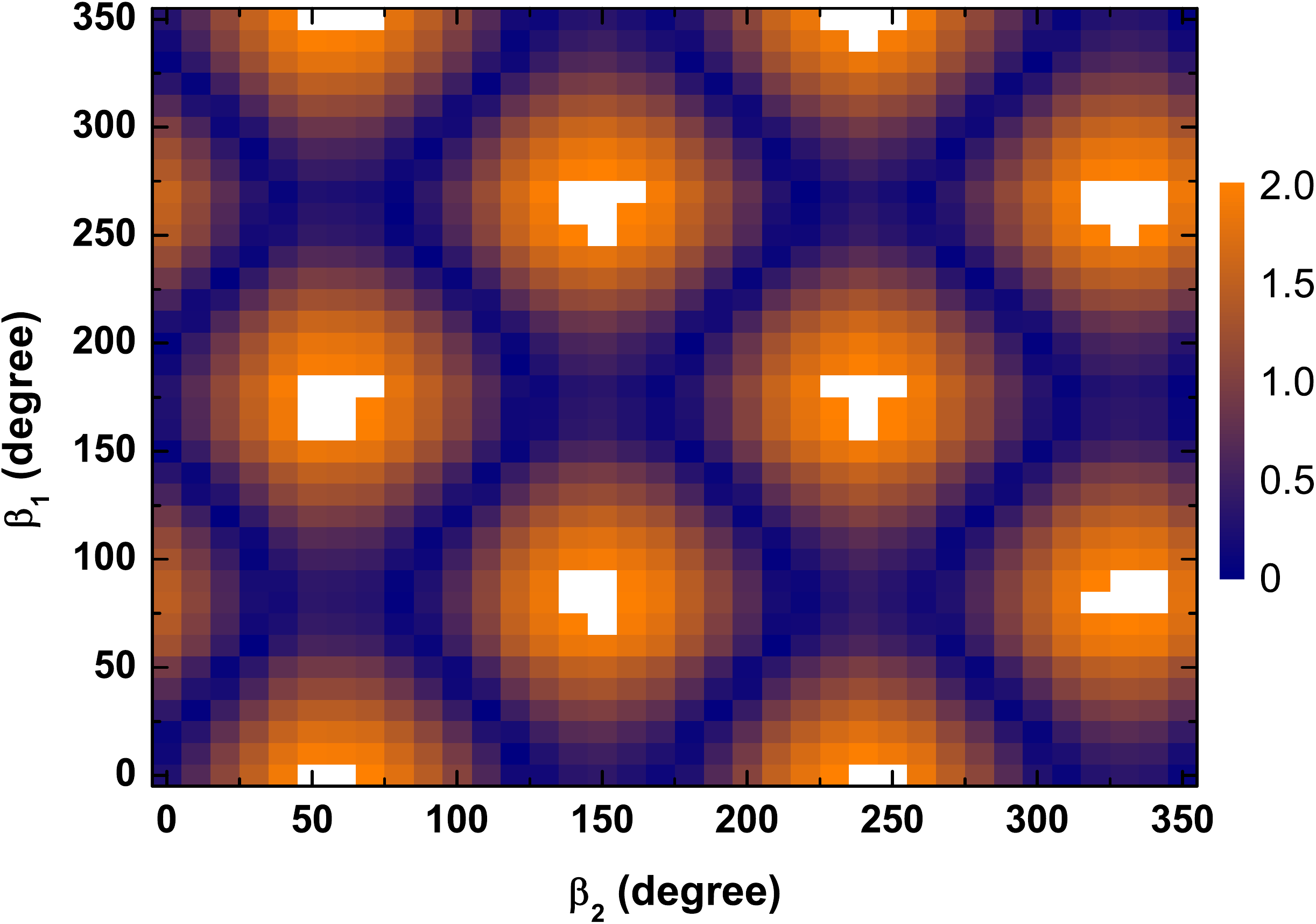}

\caption{Parameter scan of the CHSH S-parameter as a function of $(\beta_{1},\,\beta_{2})$
for fixed $(\alpha_{1},\,\alpha_{2})$. Each pixel shows color-coded
$S(\alpha_{1}=0^{\circ},\,\alpha_{2}=-45^{\circ},\beta_{1},\,\beta_{2})$,
which is calculated using 16 different coincidence count measurements
at different orientations of the phase plates. Regions where the CHSH
inequality $(S\!\leq\!2)$ is violated are colored white.}
\end{figure}

As the second step, we study the 2D subspace spanned by the \emph{non-degenerate}
states $\mid\!\! HG_{0,0}\rangle$ and $\mid\!\! HG_{1,0}\rangle$.
In this case it is necessary to consider first the intermodal dispersion
of the relevant modes in the fiber, i.e., the relative propagation
constants of the $HG_{0,0}$ and $HG_{1,0}$ modes. We measure this
by launching a well-known (classical) coherent superposition of these
modes into the fiber and observe, as a function of wavelength, the
near-field intensity at the fiber exit. From the mode beating we determine
the intermodal dispersion \cite{hlubina1995} to be 1.5~ps/m. This
agrees well with the calculated value of 1.6~ps/m obtained by approximating
the HC-PCF by a hollow dielectric capillary \cite{marcatili1964}
(with $r=12.5$~\textgreek{m}m, $\lyxmathsym{\textgreek{l}}=826$~nm,
and $n=1.45$) %
\footnote{The intermodal dispersion of our kagomé HC-PCF is accordingly much
lower than that of conventional index-guided multi-mode fibers (around
50~ps/m)%
}. Although this intermodal dispersion is small, we need to limit the
bandwidth of the down-converted photons, in order to preserve the
coherence of the non-degenerate quantum state; we use 1-nm bandwidth
filters in the experiments with the 30-cm long HC-PCF.

In this non-degenerate case we have performed a quantum interference
experiment somewhat similar to that of Mair et al. \cite{mair2001}:
In analyzer A, we no longer center the phase plate but give it a transverse
(i.e. in-plane) offset $\Delta_{PP,A}$ with respect to the single-mode
fiber (Fig. 1a). This detector configuration projects the incoming
photon onto a superposition state $a_{0}\mid\!\! HG_{0,0}\rangle+a_{1}\mid\!\! HG_{1,0}\rangle$.
The intensity distribution corresponding to this superposition carries
a nodal line. This is probed non-locally by using analyzer B as a
Gaussian probe, i.e. we remove its phase plate. For a certain offset
$\Delta_{PP,A}$, we scan the fundamental-mode analyzer B ($\Delta_{SMF,B}$)
transversely to the optical axis and normal to the edge of the phase
plate of analyzer A. We expect that the coincidence fringe resembles
the coherent superposition onto which photon A is projected, and that
the dip in this fringe moves proportionally to the phase plate A offset
$\Delta_{PP,A}$. This nonlocal shift is clearly visible in our experimental
results shown in Fig.~4: We observe a dip in the coincidence counts
(normalized to the single counts) at twice the phase step offset $\Delta_{PP,A}$,
in agreement with theory (the beam waist at the phase plates is $w_{0}=0.8$~mm).
Considering that the theory does not include mode-dependent attenuation
or intermodal dispersion of the $HG_{0,0}$ and $HG_{1,0}$ modes
in the fiber, the overall correspondence with the experimental data
is good. This demonstration of nonlocality is obviously not equivalent
to a Bell test of entanglement, but can be seen as a first step in
that direction.

\begin{figure}
\includegraphics[width=1\columnwidth]{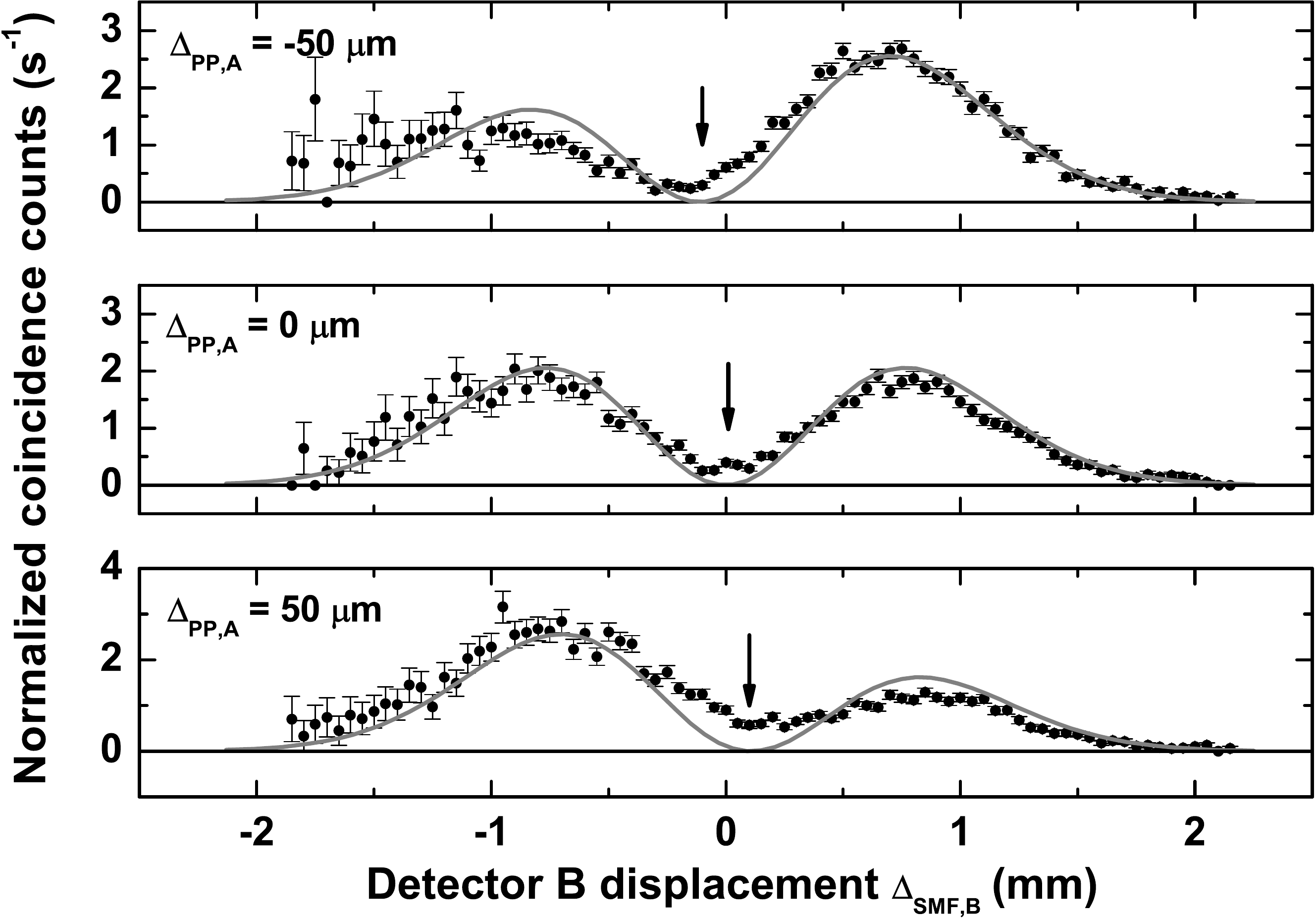}

\caption{Fiber transport of a non-degenerate superposition of photons modes.
Analyzer A projects on a superposition of $\mid\!\! HG_{0,0}\rangle$
and $\mid\!\! HG_{1,0}\rangle$, analyzer B (Gaussian probe) scans
normal to the phase plate step of analyzer A. The dip in coincidence
counts (arrow), is directly proportional to the phase plate displacement
$\Delta_{PP,A}$. The experimental coincidence count rate is normalized
to the single-detector count rate, and the theoretical curves (through
lines) are scaled vertically with the same constant to match the data.
The visibility of the three curves is 76~\%, 74~\% and 55~\% (from
top to bottom). We define the visibility here as (MAX\textendash{}MIN)/(MAX
+ MIN) where MAX is the average of the count rates in the two peaks
that are adjacent to the dip, and MIN is the count rate in the dip.}
\end{figure}

Finally, we discuss briefly the key issues in fiber transport of multi-mode
quantum superpositions: intermodal dispersion and intermodal mixing.
Intermodal dispersion leads to decoherence of the quantum superposition
during fiber transport, this must be small compared to the photon
bandwidth. Therefore, the spread in the transverse wavevector $k_{\perp}$of
the modes should be small. This is the case for weakly confining fibers
with a large core and consequently large mode area (compared to the
wavelength). Additionally, one could employ a special subset of the
modes with nearly equal propagation constants; for large-core fibers
(with negligible diffraction corrections) these are the modes within
one mode order $m+n$ in the case of $HG_{m,n}$ modes \cite{ma2009}.
However, the future applicability of fiber transport of spatially
entangled photons will be ultimately limited by unavoidable fabrication
tolerances as well as externally applied strain, stress and bends;
this does eventually lead to intermodal mixing. Due to geometric reasons
a large-diameter mode, as appearing in large-core fibers, is less
sensitive to perturbations. Since perturbations act via local refractive-index
changes, it is advantageous, as stressed in the introduction, to use
a hollow-core fiber to transport the light essentially in air, so
as to minimize these effects. Since index-guidance is impossible in
hollow-core fibers, guidance in such fibers must be preferably provided
by a photonic bandgap of the cladding. Our kagomé-lattice fiber does
not provide such lossless guiding and is therefore not ideal; mode
dependent losses may lead to non-orthogonality of the modes and thus
to mode mixing \cite{siegman1989}. It is encouraging that this potential
problem has not prevented the experiments reported in this Letter. 

In conclusion, we have shown the first experimental demonstration
of the transport of spatially entangled photons through an optical
fiber. In the degenerate subspace of the $\mid\!\! HG_{1,0}\rangle$
and $\mid\!\! HG_{0,1}\rangle$ states, we demonstrated entanglement
of the fiber-transported spatial qubit by Bell inequality violation.
Although this entanglement is demonstrated only in the context of
a degenerate qubit Hilbert space, given the known and previously demonstrated
potential of spatial modes to provide higher-dimensional entanglement
($d$>2) \cite{mair2001,vaziri2002,langford2004,osullivanhale2005}
our results can be regarded as a proof-of-principle that the transport
of higher-dimensional ($d$>2) spatially entangled photons through
fibers might be possible in future. Our demonstration of quantum interference
in a nondegenerate 2D Hilbert space (Fig.~4) is a first step in this
direction.
\begin{acknowledgments}
We acknowledge fruitful discussions with M. P. van Exter and G. Nienhuis
and financial support by NWO and the EU FET-Open grant HIDEAS (FP7-ICT-
221906).
\end{acknowledgments}

\newpage{}
\end{document}